\documentclass[a4paper,11pt]{article}
\usepackage{pos}
\usepackage{physics}
\usepackage{subcaption}
\usepackage{todonotes}

\newcommand{\CME}{C^{\rm neq}_{\rm CME}}
\newcommand{\CMEMP}{C^{\rm MP}_{\rm CME}}

\title{Out-of-equilibrium Chiral Magnetic Effect via Kubo formulas}

\author[a]{B. B. Brandt}
\author[a,b]{G. Endr\H{o}di}
\author*[a,b]{E. Garnacho-Velasco}
\author[a]{G. Mark\'{o}}
\author[a,c]{A. D. M. Valois}

\affiliation[a]{Fakult\"{a}t f\"{u}r Physik, Universit\"{a}t Bielefeld,\\
Universit\"{a}tsstra{\ss}e 25, 33615 Bielefeld, Germany}
  
\affiliation[b]{Institute of Physics and Astronomy,
ELTE E\"otv\"os Lor\'and University,\\
P\'azm\'any P.\ s\'et\'any 1/A, H-1117 Budapest, Hungary}  
  
\affiliation[c]{CAFPE and Departamento de F\'{i}sica Te\'{o}rica y del Cosmos,\\ 
Universidad de Granada, E-18071 Granada, Spain}

\emailAdd{brandt@physik.uni-bielefeld.de}
\emailAdd{gergely.endrodi@ttk.elte.hu}
\emailAdd{egarnacho@physik.uni-bielefeld.de}
\emailAdd{gmarko@physik.uni-bielefeld.de}
\emailAdd{dvalois@physik.uni-bielefeld.de}

\abstract{In this proceedings article, we present the first steps towards the determination of the out-of-equilibrium conductivity of the Chiral Magnetic Effect (CME) in the presence of strong interactions. Using linear response theory, we obtain an analytical expression for the spectral function associated with this effect at one-loop in perturbation theory. In addition, we provide a first estimate of the CME conductivity by calculating the associated Euclidean correlator using quenched Wilson fermions and dynamical staggered fermions in physical Quantum Chromodynamics (QCD) simulations. In particular, we focus on the midpoint of the correlator, which can be used as a proxy of the full conductivity. We present results in a wide range of temperatures, showing how this observable is suppressed at low temperatures, while at high temperatures it approaches the perturbation theory prediction.}

\FullConference{%
    The 41st International Symposium on Lattice Field Theory (LATTICE2024)\\
    28 July - 3 August 2024\\
    Liverpool, UK
}

\begin{document}
\maketitle

\section{Introduction}
Anomalous transport phenomena constitute one of the most interesting manifestations of the non-trivial topology of the QCD vacuum. Its appearance is a direct consequence of the interaction of electromagnetic fields or vorticities with quantum anomalies. A prominent example is the Chiral Magnetic Effect (CME), which entails the generation of an electric current in the presence of a chiral imbalance and a background magnetic field~\cite{Fukushima:2008xe}. Since its inception, the CME has been subject to a broad and intense study, not only from a theoretical perspective, but also experimentally. Although this effect has been indirectly measured in Weyl-semimetals~\cite{Li:2014bha}, a conclusive detection of the CME in heavy-ion collision experiments has remained elusive so far, see~\cite{Kharzeev:2024zzm} for a recent review. For an overview of recent theoretical investigations of CME-related observables using lattice simulations, see~\cite{Endrodi:2024cqn}.

From a theoretical point of view, the CME is currently understood as an out-of-equilibrium effect, since its equilibrium counterpart is forbidden by the generalized Bloch theorem~\cite{Yamamoto:2015fxa}, see the recent lattice results~\cite{Brandt:2024wlw}, also accounted for in~\cite{Adhikari:2024bfa}. Whereas the theorem prohibits the existence of global currents, local currents may still exist, as long as their volume average equals zero. In a previous work, we showed that an inhomogeneous magnetic field leads to a non-trivial localized signal for the CME~\cite{Brandt:2024fpc}. 

Little is known about the out-of-equilibrium conductivity of the CME in QCD. Linear response theory provides an opportunity to study this quantity from the equilibrium formulation, but most studies have only addressed this task for non-interacting fermions~\cite{Kharzeev:2009pj,Hou:2011ze,Banerjee:2022snd}. A first step in the study of interactions has been presented in Ref.~\cite{Buividovich:2024bmu}. Another possibility is to probe the CME indirectly, by the enhancement of the parallel electric conductivity at finite magnetic field. This has been observed in lattice simulations in physical QCD at high temperature~\cite{Astrakhantsev:2019zkr,Almirante:2024lqn}  and in quenched two-color QCD simulations at $T=0$~\cite{Buividovich:2010tn}. 

In this work, we aim at a direct calculation of the CME conductivity by considering a system at a finite background magnetic field that is perturbed with a time-dependent chiral chemical potential $\mu_5(t)$. Using linear response theory, we calculate the corresponding spectral function in one-loop perturbation theory, which can be related to the CME conductivity via a Kubo formula. In addition, we give first estimates of this conductivity in QCD using the midpoint of the associated correlation function~\cite{Buividovich:2024bmu}. At various temperatures, using lattice simulations with quenched Wilson fermions and dynamical staggered quarks in QCD at the physical point.  

\section{CME in linear response theory}
We consider a system in the presence of a background magnetic field, which, without loss of generality, we choose to point in the third spatial direction: $\Vec{B}=B\,\Vec{e}_3$. In this case, the CME current is induced in the direction parallel to the field, hence we focus on the vector current $J_3$. The chiral imbalance, required for the existence of the CME, is parameterized by a chiral chemical potential $\mu_5$. Within linear response theory, we consider a time-dependent perturbation $\delta\mu_5(t)$ at nonzero background magnetic field $B$. The response of the system in terms of the CME is characterized by the axial-vector retarded propagator in Minkowski space $G_{R}$,
\begin{equation}
    \label{eq:retCMEdef}
    G^{\rm CME}_{R}(t)=i\theta(t) \dfrac{1}{V}\expval{[J_{45}(t),J_3(0)]}\,,
\end{equation}
with the (charged) vector current and the (baryon) axial currents defined as 
\begin{align}
    J_\nu(t)&=\sum_f\qty(\dfrac{q_f}{e})\int \dd^3x \, \bar{\psi}_f(x)\gamma_\mu\psi_f(x)\,,\\
    J_{\nu5}(t)&=\sum_f \int \dd^3x \, \bar{\psi}_f(x)\gamma_\mu\gamma_5\psi_f(x)\,,
\end{align}
where the sum runs over the number of flavors in the system and $q_f$ is the associated electric charge of each flavor.

Using a Kubo formula, we can relate the out-of-equilibrium CME conductivity $\CME$ to the spectral function $\rho_{\rm CME}$, 
\begin{equation}
  \CME\equiv\lim_{B\rightarrow0}\dfrac{T}{eB\,C_{\rm dof}}  \lim_{\omega\rightarrow0}\dfrac{\rho_{\rm CME}(\omega)}{\omega}\,,
    \label{eq:kubo}
\end{equation}
with
\begin{equation}
    \rho_{\rm CME}(\omega) =\frac{1}{\pi}\,\Im\widetilde{G}_{R}(\omega)\,.
\end{equation}

Here $C_{\rm dof}$ is an overall proportionality factor that depends on the details of the system, like the number of colors $N_c$, the number of flavors and its electric charges, in particular $C_{\rm dof}=N_c\sum_f (q_f/e)^2$. 

The one-loop contribution to the spectral function is a pure Quantum Electrodynamics (QED) diagram, which does not involve any gluonic interactions. We can evaluate it by an analytical continuation of the Euclidean counterpart of Eq.~\eqref{eq:retCMEdef}, using free fermion propagators in a magnetic background. The result for a single fermion of unit charge is given by 
\begin{align}
\label{eq:rho_CME}
    &\frac{1}{eB}\rho_{\rm CME}(\omega)=\alpha(m/T)\omega\delta(\omega)+\Theta(\omega^2-4m^2)\dfrac{m^2}{\pi^2}\dfrac{\tanh[|\omega|/(4T)]}{\omega\sqrt{\omega^2-4m^2}}\,,
\end{align}
with
\begin{equation}
   \alpha(m/T)= -\dfrac{1}{\pi^2}\int_0^\infty \dd p\, \dfrac{p^2}{E_p^2}n_F^{\prime}(E_p)\xrightarrow{m/T\rightarrow0}\dfrac{1}{2\pi^2}\,,
\end{equation}
with $E_p=\sqrt{p^2+m^2}$ and $n_F^{\prime}(E_p)$ the derivative of the Fermi-Dirac distribution. Notice that the conductivity defined by Eq.~\eqref{eq:kubo} diverges at this order, a common issue also present in the calculation of the electric conductivity, which is solved by the resummation of higher orders in the perturbative series~\cite{Arnold:2000dr}. 

There is an interesting comparison that can be drawn: we can consider another anomalous transport effect, the Chiral Separation Effect (CSE), the generation of an axial current in a dense and magnetized system~\cite{Son:2004tq,Metlitski:2005pr}. Analogously to the CME case, we can use linear response theory (exchanging the chiral chemical potential perturbation for one with a usual chemical potential $\delta\mu(t)$), so that we can define the retarded propagator,
\begin{equation}
\label{eq:Gr_CSE}
    G^{\rm CSE}_{R}(t)=i\theta(t) \dfrac{1}{V}\expval{[J_{4}(t),J_{35}(0)]}\,,
\end{equation}
and calculate its associated spectral function (again using one-loop perturbation theory),
\begin{align}
    \label{eq:rho_CSE}
    &\frac{1}{eB}\rho_{\rm CSE}(\omega)=\alpha(m/T)\omega\delta(\omega)\,.
\end{align} 
At this order in the perturbative expansion, the spectral function of the CSE is only non-zero at $\omega=0$, where it is equal to the spectral function of the CME. This also implies that the Euclidean CSE propagator is $\tau$-independent, where $\tau$ is the Wick-rotated imaginary time. In addition, this means that both the CME and CSE Euclidean correlators are equal at the midpoint $\tau=1/(2T)$, a property that will play an important role below. 

Lattice simulations do not have direct access to real-time quantities such as the retarded propagator or the spectral function, since the theory is formulated in the imaginary time formalism. However, Euclidean propagators, which are easily obtainable on the lattice, can be related to the spectral function $\rho$ using the following spectral representation
\begin{equation}
    G(\tau)=\int_0^\infty \dd \omega \, \dfrac{\rho(\omega)}{\omega}\,K(\tau,\omega)\,,\hspace{1cm}K(\tau,\omega)=\dfrac{\omega\, \cosh[\omega(\tau-1/(2T))]}{\sinh[\omega/(2T)]}\,.
    \label{eq:spec_rep}
\end{equation}
The extraction of the spectral function from the Euclidean correlator is a well-known inverse problem, defined by Eq.~\eqref{eq:spec_rep}. Since the correlator $G(\tau)$ can only be calculated in $N_t$ discrete points, this is an ill-posed problem, and a precise extraction of the conductivity is a challenging task. Although there are many methods available in the literature to attempt to diminish its issues, in this work we merely take a first step in the study of the CME conductivity, following Ref.~\cite{Buividovich:2024bmu}. In particular, we consider the correlator evaluated at the midpoint $G(\tau=1/(2T))$, which carries a first estimate of this conductivity. This can be seen using the spectral representation~\eqref{eq:spec_rep}, since the kernel $K(\tau,\omega)$ evaluated at $\tau=1/(2T)$ approaches a Dirac delta $\delta(\omega)$ in the limit $T\rightarrow0$. Hence we define the observable 
\begin{equation}
    \CMEMP=\dfrac{1}{C_{\rm dof}}\dfrac{G\qty(\tau T=1/2)}{eB\,T}\,,
\end{equation}
where we have used the factor $T^2$ to normalize the observable, since
\begin{equation}
    \int_0^\infty \dd \omega\, K(1/(2T),\omega)\propto T^2\,.
\end{equation}
Using Eq.~\eqref{eq:rho_CME}, the analytical value of this observable in the free case can also be calculated,
\begin{equation}
\label{eq:MP anly}
     \CMEMP(m/T) =
     \frac{1}{2\pi^2} \int_{0}^{\infty} \dd p \, \left[ 1+\cosh( \sqrt{p^2+(m/T)^2} ) \right]^{-1}\,.
\end{equation}
We note that this expression is equal to the value of the equilibrium CSE conductivity~\cite{Brandt:2023wgf}, which is a consequence of the form of the CSE spectral function~\eqref{eq:rho_CSE} and the $\tau$-independence of the CSE correlator, as we discussed above.

\section{Lattice setup}

We use two different fermion discretizations in our lattice simulations, in particular staggered and Wilson fermions. Here, we introduce the general definition of the Euclidean CME correlator on the lattice, which will be useful to define the specific form of this correlator in each discretization later on. This correlator is defined as follows
\begin{equation}
G(\tau-\tau^{\prime}) = \dfrac{1}{V}\fdv{\expval{J_3(\tau)}}{\mu_5(\tau^{\prime})} = \dfrac{1}{V}\expval{J_3(\tau)J_{45}(\tau^{\prime})}\,.
\label{eq:deriv_j3_wrt_mu5}
\end{equation}
This expectation value contains connected, disconnected, and tadpole terms, which read
\begin{align}
G_{\rm conn}(\tau-\tau^{\prime}) &= -\frac{1}{V}\sum_f \frac{q_f}{e}\expval{\Tr\qty[\Gamma^f_{3} M_f^{-1}(\tau)\Gamma^f_{45} M_f^{-1}(\tau^{\prime})]}\,,\label{eq:conn_part}\\
G_{\rm disc}(\tau-\tau^{\prime}) &= \dfrac{1}{V}\sum_{f,f'}\frac{q_f}{e}\expval{\Tr\qty[\Gamma^{f}_{3} M_{f}^{-1}(\tau)]\Tr\qty[\Gamma^{f'}_{45} M_{f'}^{-1}(\tau^{\prime})]}\,, \\
G_{\rm tad}(\tau-\tau^{\prime}) &= \frac{1}{V}\sum_f  \frac{q_f}{e}\expval{\Tr\qty[\delta_{\tau,\tau^{\prime}}\pdv{\Gamma^f_{3}}{\mu_{5}}M_f^{-1}]}\,,
\label{eq:conn_disc_tadpole}
\end{align}
where $\Gamma_3$ and $\Gamma_{45}$ will be defined for staggered and Wilson fermions below. Notice that these expectation values are to be evaluated at nonzero magnetic field but vanishing chiral chemical potential $\mu_5$.

\subsection{Staggered fermions}
We consider 2+1 flavors of rooted staggered fermions at the physical point with two stout-smearing steps, and the tree-level Symanzik-improved gauge action. The staggered Dirac matrices are defined as
\begin{align}
  \label{eq:gammas}
    \Gamma^f_\nu(n,m)&=\dfrac{\eta_\nu(n)}{2}\left[U_\nu(n)u^f_{\nu}(n) \,e^{h(\mu_5)}\delta_{n+\hat{\nu},m}+U^{\dagger}_\nu(n-\hat{\nu})u^{f*}_{\nu}(n-\hat{\nu}) \,e^{-h(\mu_5)}\delta_{n-\hat{\nu},m}\right], \nonumber \\
        \Gamma^f_{\nu5} &=\dfrac{1}{3!}\sum_{\rho,\alpha,\beta} \epsilon_{\nu\rho\alpha\beta}\,\Gamma^f_\rho\Gamma^f_\alpha\Gamma^f_\beta\,,
\end{align}
where $u_{\nu}^f$ are U(1) phases, corresponding to a uniform magnetic field in the third spatial direction (see Ref.~\cite{Bali:2011qj} for the explicit form of the links). Moreover, we defined
\begin{equation}
\label{eq:hmu5}
    h(\mu_5)=a\mu_{5}\,\Sigma^f_\nu(\mu_5)\,,
\end{equation}
and the staggered representation of the spin operator,
\begin{align}
    &\Sigma^f_\nu(\mu_5)=\frac{1}{3!}\epsilon_{\nu\rho\alpha4}\Gamma^f_\rho\Gamma^f_\alpha(\mu_5)\,.
\end{align}
From Eqs.~\eqref{eq:conn_part}-\eqref{eq:conn_disc_tadpole}, the CME correlator for staggered fermions reads
\begin{equation}
G_{\rm stagg}(\tau) = \frac{1}{16}G_{\rm conn}(\tau)  + \frac{1}{4}G_{\rm tad}(\tau)\,,
\end{equation}
where the pre-factors appear due to the fourth-rooting procedure. For a further discussion on the operators, see Ref.~\cite{Brandt:2024wlw}. We have not included the disconnected part, since it was found to be significantly more noisy than the other two terms, especially at lower temperatures. This is the same approach taken in Ref.~\cite{Buividovich:2024bmu} to study $\CMEMP$ in two-color QCD with Wilson fermions. A precise analysis of the disconnected contribution will be carried out and presented in an upcoming publication.

The temporal correlators in the staggered formulation have two distinct contributions for even and odd points due to the taste doubling. The two contributions have different continuum limits, and their average yields the physical observable, see e.g.\ Ref.~\cite{Astrakhantsev:2019zkr}. Therefore, we define the observable in the staggered formulation as 
\begin{equation}
    \CMEMP=\dfrac{1}{C_{\rm dof}}\dfrac{1}{eB\,T}\dfrac{G(N_t/2)+G(N_t/2-1)}{2}\,.
\end{equation}

\subsection{Quenched Wilson fermions}

We also use Wilson fermions in the quenched approximation, with gauge configurations sampled from the Wilson plaquette gauge action. In this formulation, the conserved vector and anomalous axial current are defined via the operators
\begin{align}
         \Gamma^f_\nu(n,m)&=\frac{1}{2}\Big[(\gamma_\nu e^{a\mu_{5}\delta_{\nu4}}-r)U_\nu(n) u_{f\nu}(n) \, \delta_{m,n+\hat{\nu}} \nonumber\\
         &\qquad\quad+(\gamma_\nu e^{-a\mu_{5}\delta_{\nu4}}+r)U^\dagger_\nu(n-\hat{\nu})u^*_{f\nu}(n-\hat{\nu})\, \delta_{m,n-\hat{\nu}}\Big]\,,\\
            \Gamma^f_{\nu5}(n,m)&=\frac{1}{2}\Big[\gamma_\nu\gamma_5 U_\nu(n) u_{f\nu}(n) \,e^{a\mu_{5}\delta_{\nu4}} \delta_{m,n+\hat{\nu}}\nonumber\\
	&\qquad\quad+\gamma_\nu\gamma_5 U^\dagger_\nu(n-\hat{\nu}) u^*_{f\nu}(n-\hat{\nu})\,e^{-a\mu_{5}\delta_{\nu4}}\delta_{m,n-\hat{\nu}}\Big]\,.
\end{align}
In the Wilson case, the CME correlator reads
\begin{equation}
G_{\rm Wilson}(\tau) = G_{\rm conn}(\tau)\,. 
\end{equation}
Notice that the tadpole term vanishes in the Wilson formulation since $\Gamma_3$ does not depend on $\mu_5$. As in the staggered case, the disconnected part has not been included in the calculation.

\section{Results}

\begin{figure}[t]
    \centering
    \includegraphics[width=\linewidth]{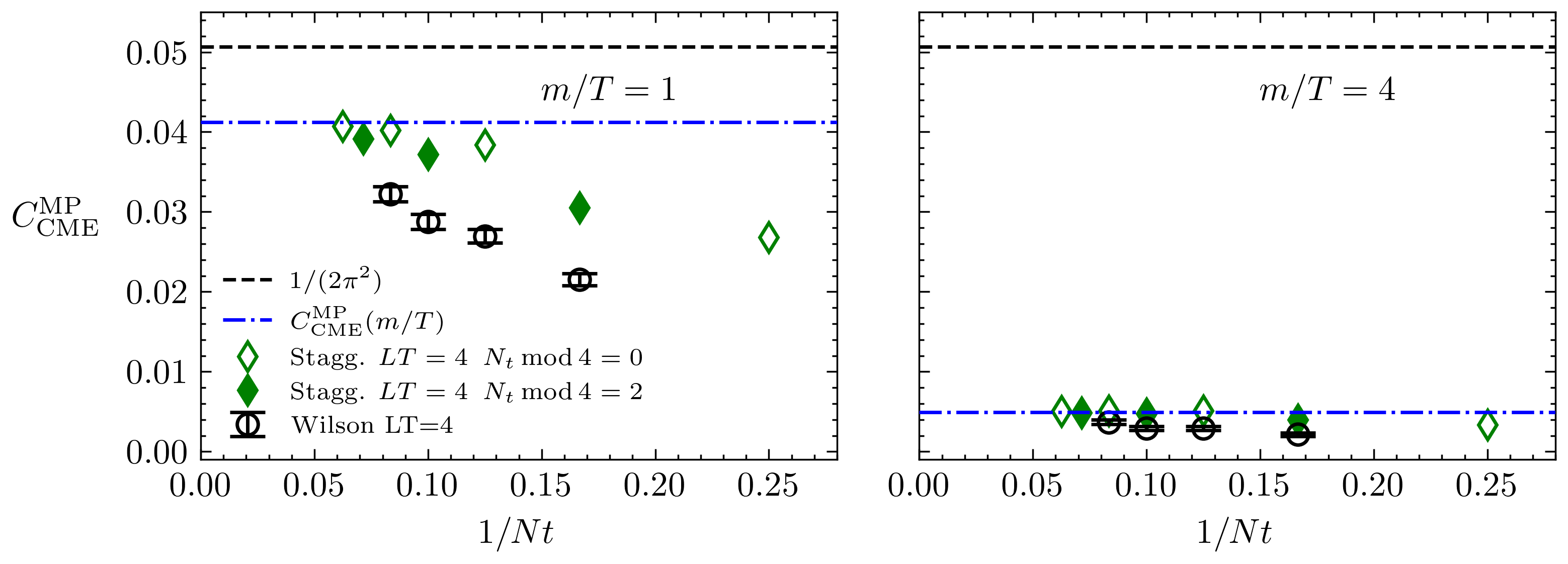}
    \caption{Left: Results for $\CMEMP$ at $m/T=1$ in the free case using Wilson and staggered fermions. Right: analogous plot at $m/T = 4$. The blue dot-dashed lines show the analytical result for the respective value of $m/T$, while the black dashed line indicates the $m/T\to0$ limit.}
    \label{fig:CME_MP_FREE}
\end{figure}

Next, we present the results obtained using our lattice simulations. First, we use non-interacting fermions to crosscheck our setup. In Fig.~\ref{fig:CME_MP_FREE}, we show the obtained results for $\CMEMP$ using staggered and Wilson fermions at two different values of $m/T$. The aspect ratio $LT=4$ was found to be sufficiently close to the thermodynamic limit. In both cases, the continuum limit approaches the analytic result. This behavior is also reproduced for other values of $m/T$. We note that in the staggered case, the approach to the continuum limit is observed to be non-monotonous -- in particular, lattices with $N_t\,\,\rm {mod}\,4=0$ are found to scale faster towards the continuum as lattices with $N_t\,\,\rm{mod}\,4=2$. We will get back to this observation when we discuss the results in QCD. In addition, we checked that our results in the Wilson formulation coincide with the free fermion results reported in Ref.~\cite{Buividovich:2024bmu}.

\begin{figure}[t]
    \centering
    \includegraphics[width=\linewidth]{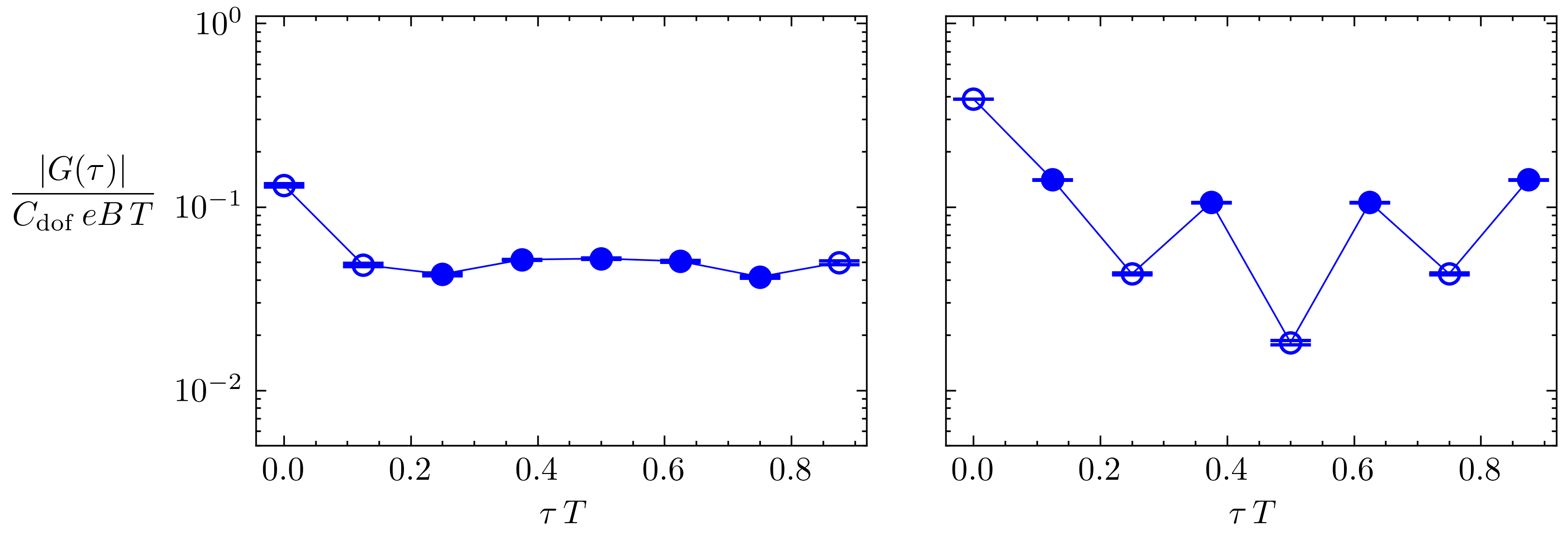}
    \caption{CME correlator for quenched Wilson fermions (left) and staggered fermions in QCD (right) on a $24^3\times8$ lattice at $T\approx300$ MeV. The (filled) open points correspond to the (positive) negative values of the correlator. In the staggered case, the contributions from the two taste partners are clearly visible.}
    \label{fig:corrs}
\end{figure}

\begin{figure}[t]
    \centering
    \includegraphics[width=\linewidth]{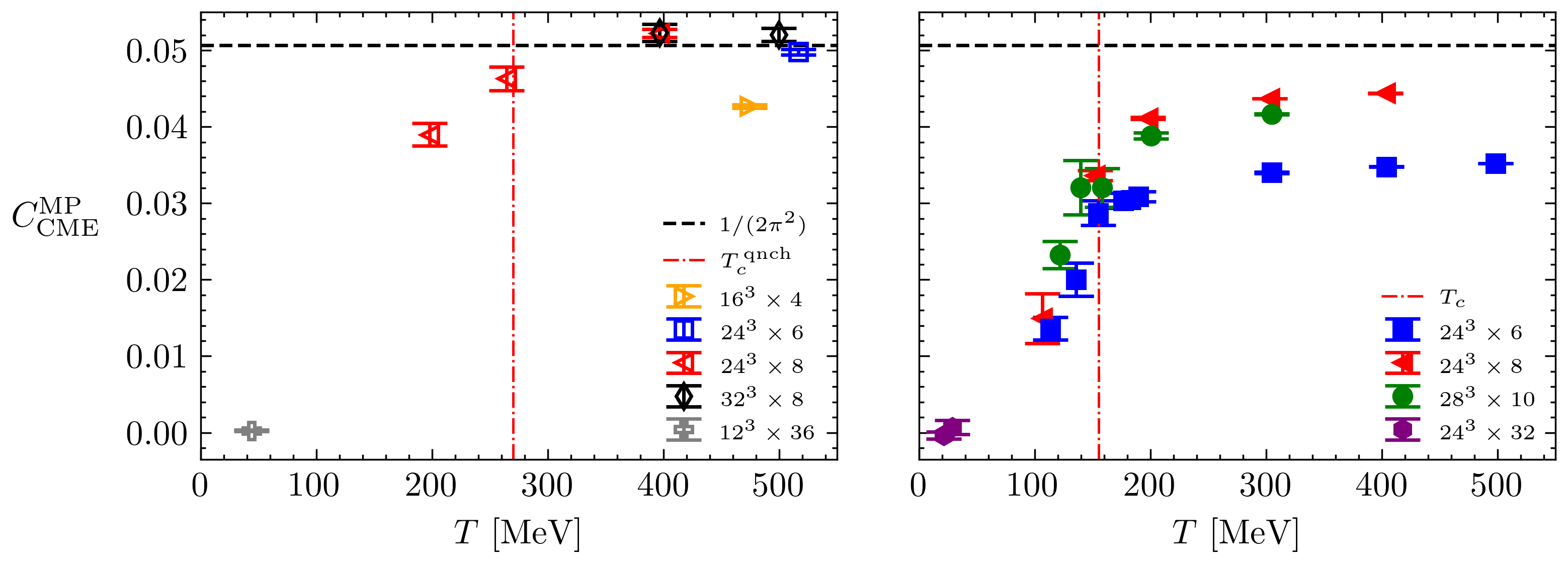}
    \caption{Value of $\CMEMP$ in QCD using quenched Wilson with $m_\pi\approx750$ MeV (left) and dynamical staggered fermions with physical quark masses (right). For comparison, the horizontal dashed line at $1/(2\pi^2)$ marks the analytical expectation in the limit $T\rightarrow\infty$. For a discussion on the lattice artifacts in the staggered formulation, see the main text.}
    \label{fig:CME_MP_QCD}
\end{figure}

Having validated our setup, we present the results for $\CMEMP$ in QCD, using quenched Wilson fermions with $m_\pi\approx750$ MeV and full dynamical simulations with staggered fermions at the physical point. In Fig.~\ref{fig:corrs}, we show the CME correlator in the presence of gluonic interactions with the two different discretizations. The results for $\CMEMP$ are shown in Fig.~\ref{fig:CME_MP_QCD}. This observable exhibits a clear suppression at temperatures below the QCD crossover transition temperature $T_c=155$ MeV, while it reveals an increase at higher temperatures, approaching the result for massless free fermions. This behavior is also reproduced in the quenched Wilson simulations, where the suppression of $\CMEMP$ begins around the quenched transition temperature $T_c^{\rm qnch}\approx 270$ MeV. We emphasize that in the staggered case, lattice artifacts may differ for the $N_t=6,10$ and $N_t=8$, as we showed for free fermions in Fig.~\ref{fig:CME_MP_FREE}. The suppression at temperatures below $T_c$ is a novel observation, which disagrees with the conclusions of Ref.~\cite{Buividovich:2024bmu} using two-color QCD simulations.

\section{Conclusions}

In this work, we studied the out-of-equilibrium conductivity of the CME using linear response theory. By considering a magnetized system perturbed with a time-dependent chemical potential, we calculated this conductivity via a Kubo formula and determined the associated spectral function within one-loop perturbation theory. In addition, we connected this spectral function with the one obtained in the case of the CSE, highlighting their similarity and the consequences of this finding for the Euclidean correlators of both effects. In particular, we emphasize that both Euclidean correlators are equal at the midpoint at this order in perturbation theory.

Using lattice simulations, we studied the midpoint $\CMEMP$ of the Euclidean CME correlator as a first step towards the complete investigation of the out-of-equilibrium properties of the CME in QCD, for instance, via spectral reconstruction techniques, which will be carried out in the future. In this work, we found that $\CMEMP$ is severely suppressed in the presence of gluonic interactions at low temperatures. Moreover, this observable is sensitive to the QCD crossover temperature $T_c=155$ MeV, exhibiting a pronounced increase in its vicinity until approaching the one-loop perturbation theory prediction. This behavior closely resembles the one found for the equilibrium CSE conductivity in QCD~\cite{Brandt:2023wgf}, hence we conjecture that the equality of the CME and CSE correlators at the midpoint may hold even in this non-perturbative system, although a more thorough analysis, involving the disconnected terms and the lattice artifacts, is required in order to establish a more precise comparison.

\acknowledgments
This work was funded by the DFG (Collaborative Research Center CRC-TR 211 ``Strong-interaction matter under
extreme conditions'' - project number 315477589 - TRR 211) and by the Helmholtz Graduate School for Hadron and Ion Research (HGS-HIRe for FAIR), as well as by STRONG-2020 ``The strong interaction at the frontier of knowledge: fundamental research and applications'' which received funding from the European Union's Horizon 2020 research and innovation programme under grant agreement No 824093. GE also acknowledges funding from the Hungarian National Research, Development and Innovation Office (Research Grant Hungary 150241) and the European Research Council (Consolidator Grant 101125637 CoStaMM).
The authors are grateful for enlightening discussions with Pavel Buividovich and Francesco Becattini.

\bibliographystyle{utphys}
\bibliography{biblio}

\end{document}